\documentclass[12pt]{article}

\usepackage{xpatch}

\usepackage{scicite}
\usepackage{graphics, graphicx}
\usepackage{times}
\usepackage{amsmath, amssymb}

\usepackage{hyperref}
\usepackage{textcomp}
\usepackage{color, soul, textcomp}

\usepackage{graphicx}
\usepackage{dcolumn}
\usepackage{bm}
\usepackage{epsfig}
\usepackage{color}
\usepackage{amsmath}
\usepackage{hyperref}
\usepackage[english]{babel}

\hypersetup{colorlinks=true, urlcolor=blue, citecolor=blue, linkcolor=blue, anchorcolor=blue}  

\newenvironment{sciabstract}{%
	\begin{quote} \bf}
	{\end{quote}}

\newcommand{\ket}[1]{\ensuremath{| #1 \rangle}}
\newcommand{\bra}[1]{\ensuremath{\langle #1 |}}

\newcommand{\be}{\begin{equation}}
\newcommand{\ee}{\end{equation}}
\newcommand{\ba}{\begin{align}}
\newcommand{\ea}{\end{align}}

\newcommand{\si}{\ensuremath{s_{\rm in}}}
\newcommand{\car}{\downarrow}
\newcommand{\cl}{\uparrow}
\newcommand{\textred}[1]{\textcolor{black}{#1}}

\title{{\bf A single photonic cavity with two independent physical synthetic dimensions}}
\author{Avik Dutt$^1$, Qian Lin$^2$, Luqi Yuan$^{3,\dagger}$, 
	Momchil Minkov$^1$, Meng Xiao$^4$, Shanhui Fan$^{1,\dagger}$ \\
	\\
	\normalsize{$^1$Ginzton Laboratory and Department of Electrical Engineering, Stanford University,}\\ \normalsize{Stanford, CA 94305, USA}\\
	\normalsize{$^2$ Department of Applied Physics, Stanford University, Stanford, CA 94305, USA}\\
	\normalsize{$^3$ State Key Laboratory of Advanced Optical Communication Systems and Networks,} \\
		\normalsize{School of Physics and Astronomy, Shanghai Jiao Tong University, Shanghai 200240, China}\\
	\normalsize{$^4$ Key Laboratory of Artificial Micro- and Nano-structures of Ministry of Education and}\\
	\normalsize{  School of Physics and Technology, Wuhan University, Wuhan 430072, China}\\
	\normalsize{$\dagger$Corresponding author. Email: yuanluqi@sjtu.edu.cn, shanhui@stanford.edu}
}

\date{}

\begin{document}
\baselineskip24pt

\maketitle

\begin{sciabstract}
\noindent
The concept of synthetic dimensions, which has enabled the study of higher-dimensional physics on lower-dimensional physical structures, has generated significant recent interest in many branches of science ranging from ultracold-atomic physics to photonics, since such a concept provides a versatile platform for realizing effective gauge potentials and novel topological physics. Previous experiments demonstrating this concept have augmented the real-space dimensionality by one additional physical synthetic dimension.  
Here we endow a single ring resonator with two independent physical synthetic dimensions. Our system  consists of a temporally modulated ring resonator with spatial coupling between the clockwise and counterclockwise modes, creating a synthetic Hall ladder along the frequency and pseudospin degrees of freedom for photons propagating in the ring. We experimentally observe a wide variety of rich physics, including effective spin-orbit coupling, magnetic fields, spin-momentum locking, a Meissner-to-vortex phase transition, and chiral currents, completely in synthetic dimensions. Our experiments demonstrate that higher-dimensional physics can be studied in simple systems by leveraging the concept of multiple simultaneous synthetic dimensions.

\end{sciabstract}

\maketitle

Recent years have witnessed a surge in interest in creating synthetic dimensions to study classical and quantum dynamics~\cite{boada_quantum_2012} in systems with extra dimensions beyond their real-space geometric dimensionality~\cite{yuan_synthetic_2018-1}. Synthetic dimensions can be formed by coupling atomic or photonic states with different internal degrees of freedom to form a lattice. These degrees of freedom could be based on the frequency, spin, linear momentum, orbital angular momentum, spatial supermodes or arrival time of light pulses~\cite{ozawa_topological_2019}. Previous experiments have provided demonstrations of ($d$+1)-dimensional physics on $d$-dimensional real-space lattices by using one extra synthetic dimension, for $d=1$~\cite{mancini_observation_2015, stuhl_visualizing_2015, lustig_photonic_2019} or $d=0$~\cite{bell_spectral_2017, dutt_experimental_2019-1, regensburger_paritytime_2012}. While theoretical proposals exist for creating two or more separate synthetic dimensions~\cite{yuan_photonic_2019, martin_topological_2017, yuan_synthetic_2018}, 
such proposals have eluded experimental observation so far. The realization of two or more synthetic dimensions is of paramount importance, since such a realization drastically simplifies the experimental requirements for studying a rich set of topologically nontrivial phenomena, e.g. the high-dimensional quantum Hall effect~\cite{klitzing_new_1980, lohse_exploring_2018, zilberberg_photonic_2018}, without needing complex higher-dimensional structures in real space.

\begin{figure}
\begin{center}
\includegraphics[width=.75\textwidth]{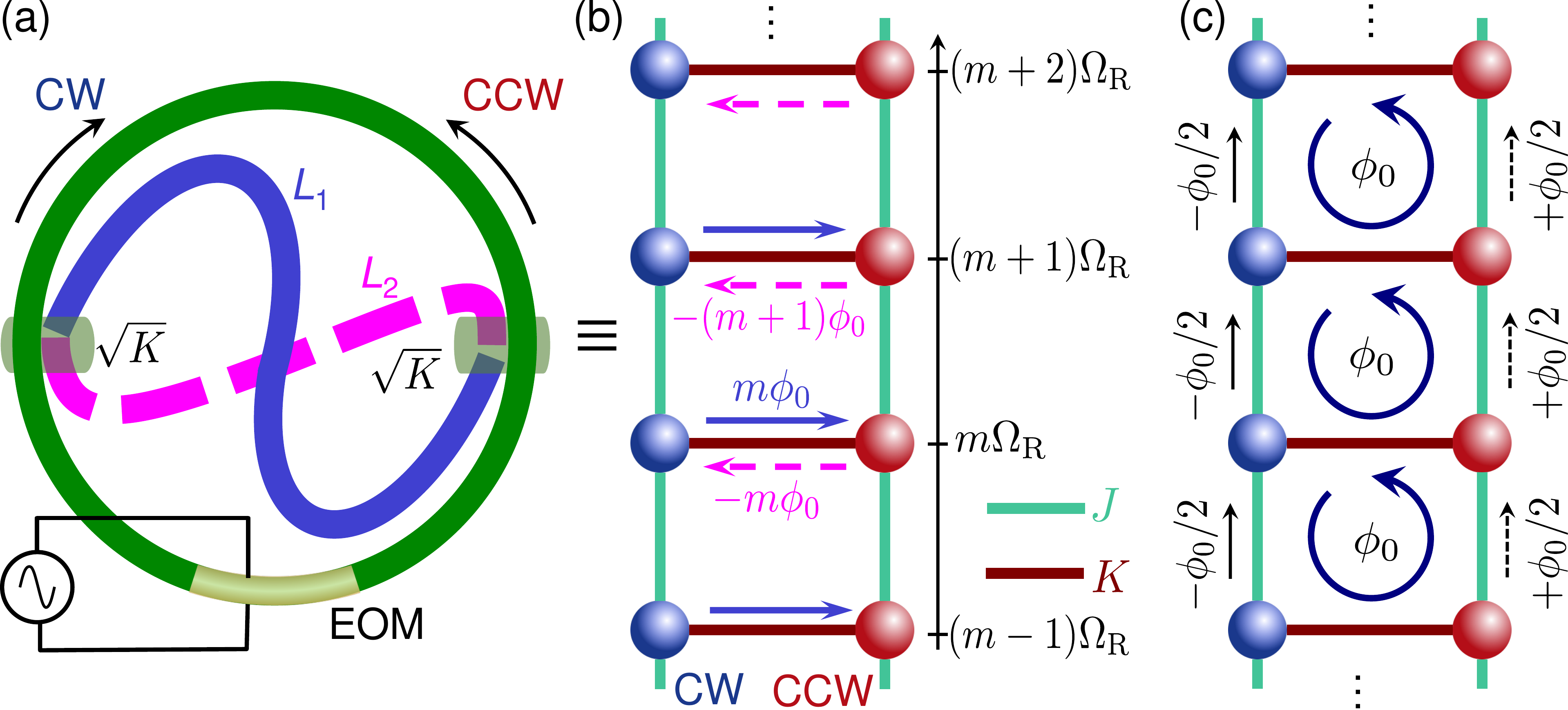}
\caption{{\bf A modulated ring resonator with clockwise-counterclockwise (CW-CCW) mode-coupling and its corresponding lattice in synthetic dimensions.} {\bf (a)} Schematic of the ring of length $L_0$ with electro-optic modulation (EOM) and CW-CCW coupling. The CW and CCW modes form the pseudospin degree of freedom. The longitudinal modes of the ring separated by the FSR $\Omega_R$ form the frequency degree of freedom. The two directional couplers are connected into an 8-shaped coupler by two connecting waveguides of unequal lengths $L_1$ and $L_2$. By varying $\Delta L = L_1 - L_2$, the phases of couplings between CW and CCW modes [in (b) and (c)] can be varied, hence realizing a controllable effective magnetic field penetrating the ladder. The corresponding synthetic lattice is shown in two equivalent gauges: {\bf (b)} A gauge with real inter-rung coupling $J$ but complex inter-leg coupling (Eq.~\eqref{eq:Hfull}), {\bf (c)} a translationally invariant gauge (Eq.~\eqref{eq:Hrwa}) with real inter-leg coupling $K$ and complex inter-rung coupling.}
\label{fig:1_schematic}
\end{center}
\end{figure}

Here we report the first demonstration of a system exhibiting two independent physical synthetic dimensions. Our system (Fig.~\ref{fig:1_schematic}(a)) consists of a ring resonator supporting a synthetic frequency dimension formed by the longitudinal cavity modes, and a synthetic pseudospin dimension formed by the clockwise (CW, $\cl$) and counterclockwise (CCW, $\car$) modes at the same frequency. The coupling along the frequency dimension is achieved with a modulator~\cite{yuan_photonic_2016}. The coupling along the pseudospin dimension is achieved with an 8-shaped coupler, consisting of two directional couplers connected by two nonintersecting waveguides. In this system, we observe a rich set of nontrivial dynamic effects, including an effective magnetic field, signatures of topological chiral one-way edge states, as well as magnetic-field controlled spin-momentum locking, entirely in the synthetic space. Demonstrations of these effects, in a pure synthetic lattice without any spatial lattice dimensions, have never been done before. 

We note that our construction is different from methods of probing higher-dimensional phenomena using topological pumps, for which the physics with two extra dimensions has been explored in recent experiments~\cite{lohse_exploring_2018, zilberberg_photonic_2018}. In these systems, a mathematical mapping between higher-dimensional lattices and lower-dimensional systems is achieved by varying some external parameters of the lower-dimensional system~\cite{yuan_synthetic_2018-1}. Although signatures of higher-dimensional physics can be observed in such topological pumping schemes, the full dynamics are not captured since the external parameters are in fact not the dynamical variables of the particles~\cite{ozawa_topological_2019}. In contrast, our approach provides the ability to explore physical dynamics in higher dimensional space.

The tight-binding Hamiltonian of our system shown in Fig.~\ref{fig:1_schematic}(a) is,
\begin{multline}
H = - \sum_{m,s} \left [ \omega_m a_{m,s}^\dagger a_{m, s} 
+   \sum_{m'}J_{mm'}(t)\, a_{m, s}^\dagger a_{m',s}\right] 
- \sum_m K a_{m,\cl}^\dagger  a_{m,\car} e^{im\phi_0} + {\rm H.c.} \label{eq:Hfull}
\end{multline}
where $a_{m,s}$ is the annihilation operator for the $m$-th longitudinal cavity mode with frequency $\omega_m = m\Omega_R$ and with pseudospin $s \in \{\car, \cl\}$. $J_{mm'}(t)$ is the coupling along the synthetic frequency dimension~\cite{yuan_photonic_2016, ozawa_synthetic_2016, bell_spectral_2017, qin_spectrum_2018}, produced by the electro-optic modulation~\cite{dutt_experimental_2019-1}. Since a small portion of the ring is modulated, this coupling can be simplified as $J_{mm'}(t) = J \cos\Omega_{\rm R}t$, i.e. the mode $m$ can couple to all the other modes of the system, and the coupling strength is independent of the mode indices~\cite{dutt_experimental_2019-1}. Here $\Omega_R$ is the free spectral range (FSR), corresponding to the separation between the longitudinal modes. $K$ in Eq.~\eqref{eq:Hfull} is the strength of the coupling between the two legs of the ladder, created by the 8-shaped coupler comprising two directional couplers with splitting amplitude $\sqrt{K}$. This coupling has a frequency-dependent and direction-dependent phase $\pm m\phi_0$ [Fig.~\ref{fig:1_schematic}(b)], with $\phi_0 = \pi \Delta L/L_0$, where $\Delta L$ is the length difference between the two connecting waveguides, and $L_0$ is the length of the ring. To explain how this phase $\pm m\phi_0$ is introduced, we note that the connecting waveguide depicted by the blue solid line in Fig.~\ref{fig:1_schematic}(a) couples exclusively from the CW to the CCW mode, whereas the connecting waveguide depicted by the dashed line couples only from the CCW to the CW mode. The phase difference between the coupling in the two directions is therefore $\Delta \phi(\omega) = \phi_{\car \rightarrow \cl} - \phi_{\cl \rightarrow \car} = \beta(\omega)\, \Delta L$, where $\beta(\omega)$ is the propagation constant at frequency $\omega$ for a mode in the connecting waveguides. Assuming that the connecting waveguides are the same as the waveguide of the ring, and since $\beta(\omega_m) = 2\pi m/L_0$, the phase difference $\Delta \phi$ increases linearly with $m$: $\Delta\phi(\omega_m) = 2\pi m\Delta L/L_0 = 2m\phi_0$. 

To transform Eq.~\eqref{eq:Hfull} into a time-independent Hamiltonian,
we define $b_{m,\cl} = a_{m,\cl} e^{-im(\Omega_R t+\phi_0/2)}$, and $b_{m, \car} = a_{m, \car} e^{-im(\Omega_R t - \phi_0/2)} $, and use the rotating-wave approximation to get:
\begin{multline}
H =  -{J\over 2}\sum_{m} (b_{m+1,\car}^\dagger b_{m,\car}e^{i\phi_0/2} + b_{m+1,\cl}^\dagger b_{m,\cl}e^{-i\phi_0/2}) 
- K \sum_m b_{m,\cl}^\dagger  b_{m,\car} + {\rm H.c.}  \label{eq:Hrwa}
\end{multline}
This Hamiltonian describes a two-legged ladder pierced by a uniform magnetic field (a Hall ladder)~\cite{fang_realizing_2012}, as each plaquette is threaded by an effective magnetic flux $\phi_0$ [see Fig~\ref{fig:1_schematic}(b), (c)]. Thus, by choosing a nonzero $\Delta L$, our structure in Fig.~\ref{fig:1_schematic}(a) naturally implements an effective magnetic field. Large magnetic fluxes spanning the entire range in $[-\pi, \pi]$ are achievable by choosing appropriate $\Delta L/L_0$. Since a purely 1D lattice does not permit magnetic field effects, our system corresponds to the simplest lattice model where the physics emerging from effective magnetic fields for photons can be observed.

Instead of describing the system in Fig.~\ref{fig:1_schematic} as a two-legged ladder threaded by a uniform magnetic field, the physics of this system can alternatively be derived in terms of magnetic-field controlled spin-orbit coupling (SOC), with the CW and CCW modes of each ring representing up and down spins. Going to the quasimomentum space ($k$-space), the Hamiltonian in Eq.~\eqref{eq:Hrwa} becomes $H = \int {\rm d}k\, {{\bf b}_k^\dagger}\, \mathcal{H}(k)\, {\bf b}_k$, with ${\bf b}_k = \sqrt{\Omega/2\pi}\, \sum_m e^{im\Omega k}(b_{m,\cl}, b_{m,\car})^T$, and
\be
{\mathcal H}(k) = -J \left[{\bf 1}_2 \cos k\Omega \cos{\phi_0\over 2} + \sigma_z \sin k\Omega  \sin{\phi_0\over 2}\right] - K\sigma_x \label{eq:Hk}
\ee
Here $\sigma_{x,z}$ are Pauli matrices. To make the SOC explicit, we recast Eq.~\eqref{eq:Hk} into the form, $
{\mathcal H}(k) = {\epsilon}(k)\cdot {\bf 1} + {\bf B}_{\rm SOC}(k) \cdot {\bm \sigma}
$,
where ${\epsilon}(k) = J\cos k\Omega \cos (\phi_0/2)$, and  ${\bf B}_{\rm SOC} = (K, 0, J\sin k\Omega \sin(\phi_0/2))$. The $z$-component of ${\bf B}_{\rm SOC}$ depends on the quasimomentum $k$, signifying spin-orbit coupling~\cite{quay_observation_2010}. The degree of SOC is controlled by the effective magnetic flux $\phi_0$. With the control of the magnetic flux, therefore, our system can exhibit a rich set of physics. Here we provide three experimental observations of such physics, all controlled by the magnetic gauge potential: spin-momentum locking in the band structure, chiral currents, and a Meissner-to-vortex phase transition.

\begin{figure}
\includegraphics[width=.9\textwidth]{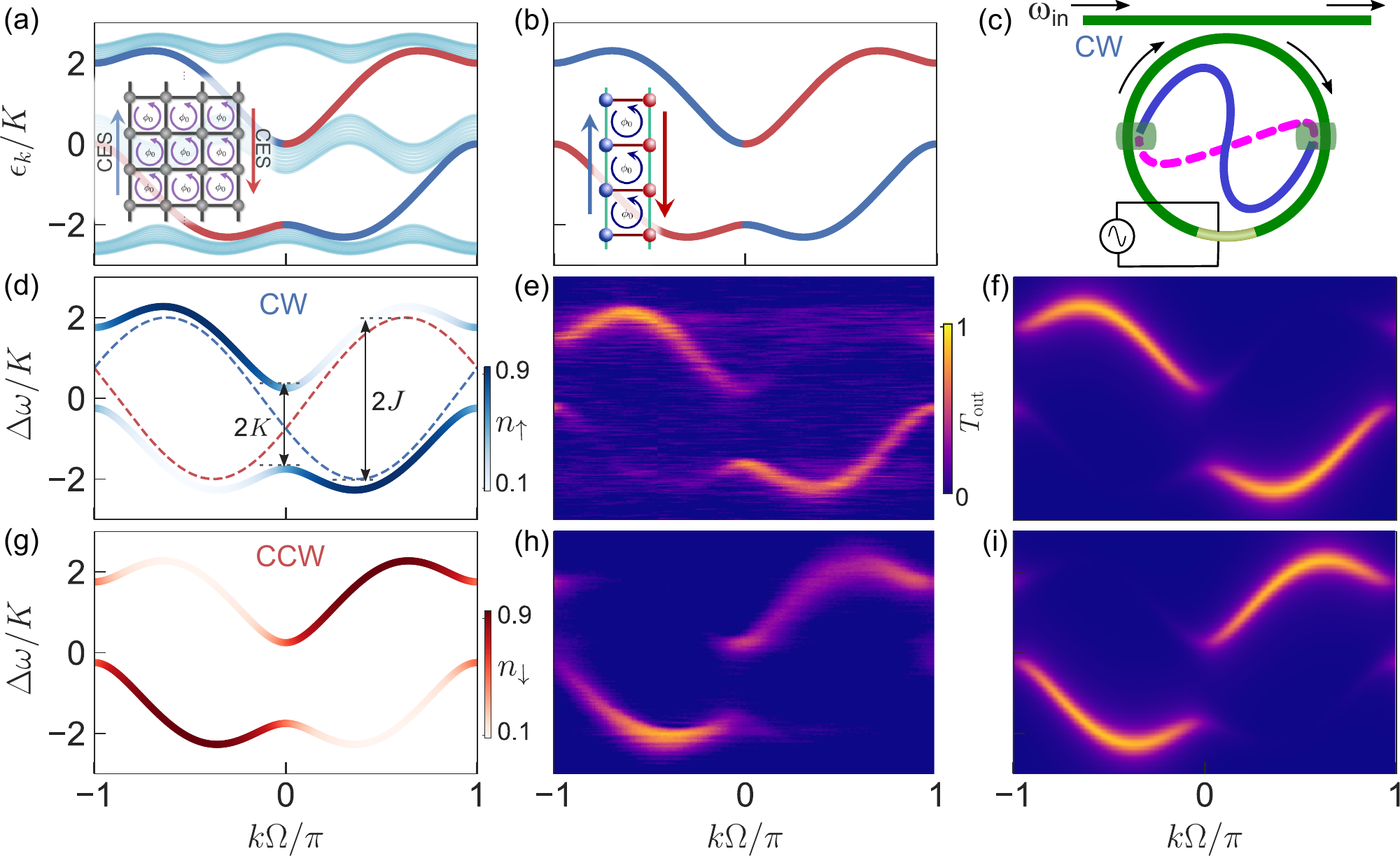}
\caption{{\bf Chiral band structure and spin-momentum locking in the synthetic Hall ladder.} {\bf (a)} Projected band structure of a 2D quantum Hall insulator (inset) infinite along $y$ and finite along $x$, showing topological chiral edge states highlighted in blue and red between the bulk band gaps. $\phi_0 = 2\pi/3$. {\bf (b)} Band structure of the two-legged synthetic Hall ladder from the tight-binding Hamiltonian $\mathcal H(k)$ (Eqs.~\eqref{eq:Hrwa} -- \eqref{eq:Hk}) for $J/K = 2$. The bulk bands disappear but signatures of chiral edge states are preserved~\cite{hugel_chiral_2014}. {\bf (c)} Schematic setup to directly measure band structure by coupling an input-output waveguide to the ring in Fig.~\ref{fig:1_schematic}(a). By varying $\omega_{\rm in}$ and detecting the time-resolved transmission through the ring, the band structure can be directly read out in experiments. The CW (CCW)- spin-resolved band structure can be detected by exciting the waveguide from the left (right) and recording its transmission. {\bf (d), (g)} Theoretical band structures, with color-coded pseudospin projections $n_{\cl}$ and $n_{\car}$ for corresponding eigenstates. For the lower band, $+k$ states have predominantly CW pseudospin character, signifying spin-momentum locking. The dashed lines are band structures for the same $J$ but for $K=0$. {\bf (e), (h)} Experimental time-resolved transmission through the ring for CW excitation [(e)] and CCW excitation [(h)]. $\Delta\omega$ is the detuning of the input frequency $\omega_{\rm in}$ from the resonance frequency of the uncoupled CW and CCW modes. {\bf (f), (i)} Theoretical time-resolved transmission based on Floquet analysis (see Supplementary Materials) . Experimental parameters: $J/2\pi = 1.95$ MHz, $K/2\pi = 0.97$ MHz. $\phi_0 \approx 3\pi/4$. Cavity linewidth $\gamma/2\pi = 480$ kHz.
}
\label{fig:2_ladder_bs}
\end{figure}

The Hall ladder has been formally shown to exactly reproduce the energies and eigenstates of the topological chiral edge modes of a 2D quantum Hall insulator [Fig.~\ref{fig:2_ladder_bs}(a)] described by the Hofstadter model~\cite{hugel_chiral_2014}. Even if the entire bulk lattice sites are removed, the strip of plaquettes forming the ladder retains the chiral currents and spin-momentum locking, as can be seen by comparing Fig.~\ref{fig:2_ladder_bs}(b) to Fig.~\ref{fig:2_ladder_bs}(a). This attests to the remarkable topological robustness of the 2D quantum Hall insulator. Such signatures of topological chiral edge modes are evident in the theoretically calculated band structure of $\mathcal H(k)$ plotted in Fig.~\ref{fig:2_ladder_bs}(d),(g) along with the corresponding color-coded pseudospin projections $n_\cl = \cos^2 (\theta_{\rm B}/2), n_\car = \sin^2 (\theta_{\rm B}/2)$ respectively. Here $\theta_{\rm B} = \arctan[K/(J\sin k\Omega \sin (\phi_0/2))]$ represents the chiral Bloch angle of the eigenstate, and its $k$-dependence signifies chiral spin-momentum locking~\cite{kolkowitz_spinorbit-coupled_2017, hugel_chiral_2014}: in the lower band, positive- (negative)-$k$ states have predominantly CW (CCW) pseudospin character. 

To directly detect the chiral modes of the Hall ladder, we use our recently introduced time-resolved band structure spectroscopy technique~\cite{dutt_experimental_2019-1}. Here we summarize this technique briefly. Our long synthetic dimension with discrete translational symmetry is the frequency axis (Eq.~\eqref{eq:Hrwa}). Hence, the corresponding quasimomentum $k$ is conjugate to  frequency, and  is identical with time $t$. On exciting the system in Fig.~\ref{fig:1_schematic}(a) using an external waveguide coupled to the ring (Fig.~\ref{fig:2_ladder_bs}(c)), time-resolved transmission measurements provide a direct momentum-resolved readout of the band structure. By scanning the detuning of the input laser frequency from the ring's resonances, we access various energies $\epsilon$, allowing us to map out the $\epsilon-k$ diagram~\cite{dutt_experimental_2019-1}. 
Furthermore, we can selectively excite the CW or CCW psuedospin by exciting the waveguide from the left or right respectively, and measure the transmitted signal to map out the band structure projected onto the corresponding spin (see Supplementary materials). 

We plot the results of these measurements in Fig.~\ref{fig:2_ladder_bs}(e),(h), which were carried out using a setup consisting of a fiber ring with an embedded electro-optic modulator and an 8-shaped coupler. We drive the modulator at $\Omega = 2\Omega_R = 29.6$ MHz (see Supplementary Material and Ref.~\cite{dutt_experimental_2019} for details on the setup). The measured band structure [Fig.~\ref{fig:2_ladder_bs}(e),(h)] agrees with that from the tight-binding model [Fig.~\ref{fig:2_ladder_bs}(d),(g)], and also with simulations using a rigorous Floquet analysis (Fig.~\ref{fig:2_ladder_bs}(f),(i), see Supplementary Materials). This constitutes the first measurement of the dispersion of \textred{chiral one-way} 
states in synthetic dimensions. It is analogous to direct methods of interrogating surface-state dispersions in SOC topological insulators (using angle-resolved photoemission spectroscopy, ARPES)~\cite{hsieh_topological_2008, jozwiak_photoelectron_2013}, or interrogating helical edge state dispersions in real-space photonic crystals~\cite{parappurath_direct_2018}. Spin-momentum locking is clearly seen in the experimental data [Fig.~\ref{fig:2_ladder_bs}(c)], as the CW mode transmission predominantly peaks at positive quasimomenta for the lower band. Here we also observe that the direction of spin-momentum locking switches for the upper band.

\begin{figure}
\begin{center}\includegraphics[width=.75\textwidth]{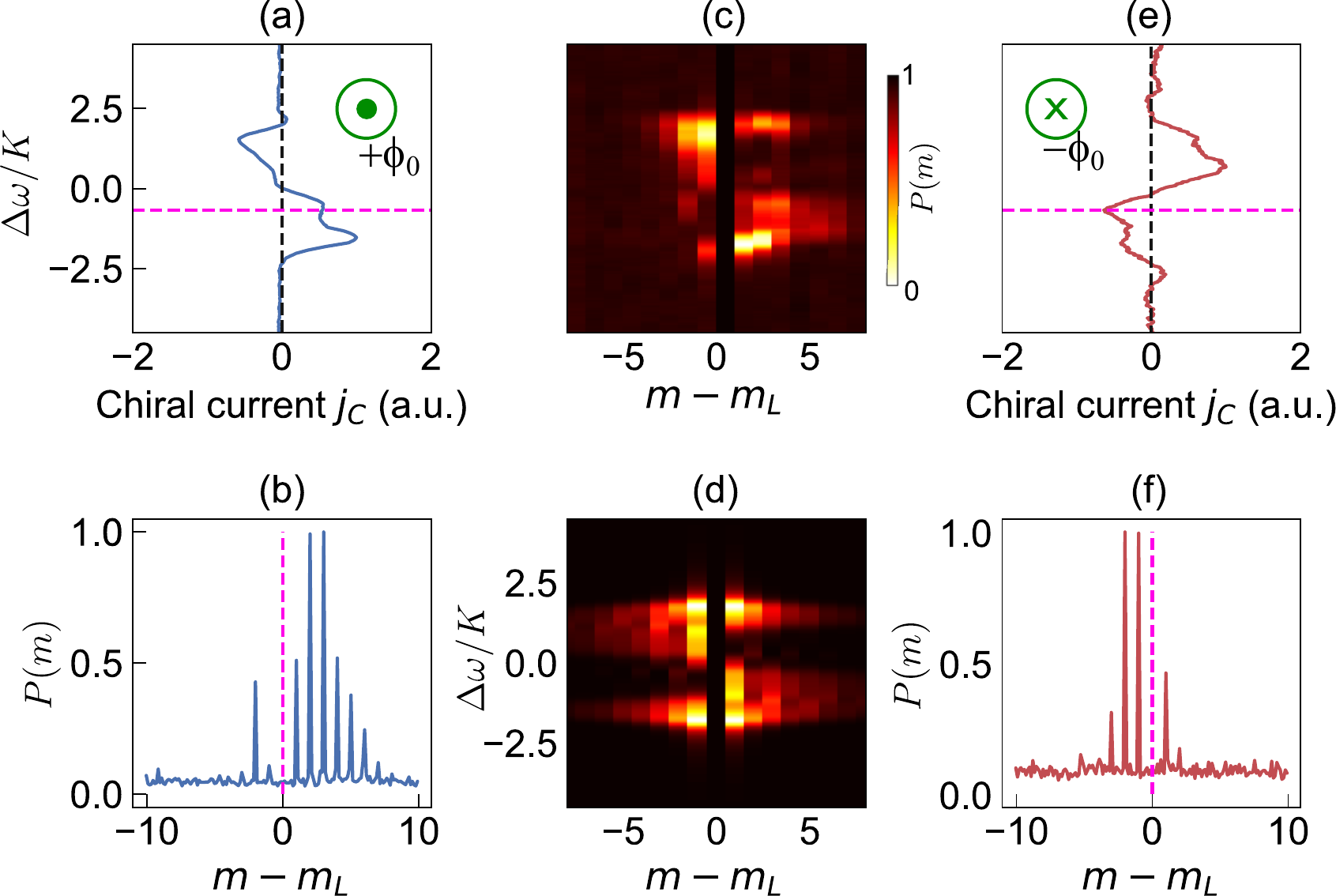}
\caption{{\bf Direct measurements of chiral currents in the synthetic Hall ladder through heterodyne detection.} {\bf (a)} Chiral current $j_{\rm C}$ (Eq. \eqref{eq:jc}) vs. laser-cavity detuning $\Delta\omega$ measured by heterodyne mixing the cavity output field with a frequency shifted part of the input laser. The full heterodyne signal is shown in (e). The lower band shows a positive $j_{\rm C}$ for the CW mode. {\bf (b)} Steady state normalized photon number of the modes at frequencies $m\Omega$ in the lower band, at $\Delta\omega/K = -0.67$ indicated by the magenta dashed line in (a). The asymmetric frequency mode occupation verifies that the CW mode predominantly evolves towards higher frequencies in the lower band. {\bf (c)} Experimental heterodyne spectra mapping out the steady state photon numbers for all $\Delta\omega$. {\bf (d)} Theoretically calculated photon numbers based on a Floquet analysis. {\bf (e), (f)} Same as in (a), (b), but with the direction of the effective magnetic field flipped, which causes a change in the sign of $j_{\rm C}$. (a) and (c) also reveal a switching of the direction of chiral current on moving from the lower to the upper band. In (c) and (d), the strong signal in the excited mode ($m-m_{\rm L}=0$) has been suppressed to reveal the occupation of other modes clearly.} 
\label{fig:4_chiral_current}
\end{center}
\end{figure}

The Hall ladder exhibits chiral currents -- in our system, the CW (CCW) pseudospin evolves preferentially to higher (lower) frequency modes for the lower band. The direction of the current switches for the upper band. To quantify direction of such spin- and band- dependent frequency evolution we define the steady-state chiral current as,
\be
j_{\rm C} = \sum_{m>m_{\rm L}} P(m, \cl) - \sum_{m<m_{\rm L}} P(m, \cl) \label{eq:jc}
\ee
where $m_{\rm L}$ is the order of the ring resonance closest to the input laser ($|\omega_{\rm in}-m_{\rm L}\Omega|<\Omega_{\rm R}/2$) and $P(m, \cl)$ is the steady-state photon number of the CW mode at frequency $m\Omega_{\rm}$.
To measure $j_{\rm C}$, we use frequency- and spin-resolved heterodyne detection of the modal photon numbers in the lattice (see Supplementary materials).  
Specifically, we frequency-shift a portion of the input laser by $\delta\omega = 500$ MHz using an acousto-optic modulator and interfere it with the cavity output. Here $\delta \omega \gg |m| \Omega$ for all the modes that we consider. A fast Fourier transform (FFT) of this interferogram directly yields $P(m)$. Heterodyne detection, i.e. the use of a frequency shift as mentioned above, is essential. If one were to set $\delta \omega = 0$ in the experiment described above, one could not distinguish between the photon numbers at $m_{\rm L} + m$ and $m_{\rm L} -m$ modes since they produce beat notes at the same radio frequency $m\Omega$. Fig.~\ref{fig:4_chiral_current}(a) shows the measured chiral current $j_{\rm C}$ versus the laser detuning $\Delta\omega$. 
For each $\Delta\omega$, $j_{\rm C}$ is calculated from the heterodyne FFT spectrum. An example of such a spectrum at $\Delta \omega /K = -0.67$ is shown in Fig.~\ref{fig:4_chiral_current}(b). In Fig.~\ref{fig:4_chiral_current}(c), we show such spectra for all $\Delta \omega$. In Fig.~\ref{fig:4_chiral_current}(d), we show a theoretical computation of the same spectrum. The overall shape of the theoretical spectrum agrees with the experiments. In both the theory and experimental results, in the lower band, the higher frequency modes have a larger occupation ($j_{\rm C}>0$). The sign of $j_{\rm C}$ is switched for the upper band. Alternately, the sign of $j_{\rm C}$ can be switched by changing the direction of the effective magnetic field [Fig.~\ref{fig:4_chiral_current}(e), (f)], which corresponds to exchanging the lengths $L_1$ and $L_2$ in our system in Fig.~\ref{fig:1_schematic}(a).


\begin{figure}
\includegraphics[width=\textwidth]{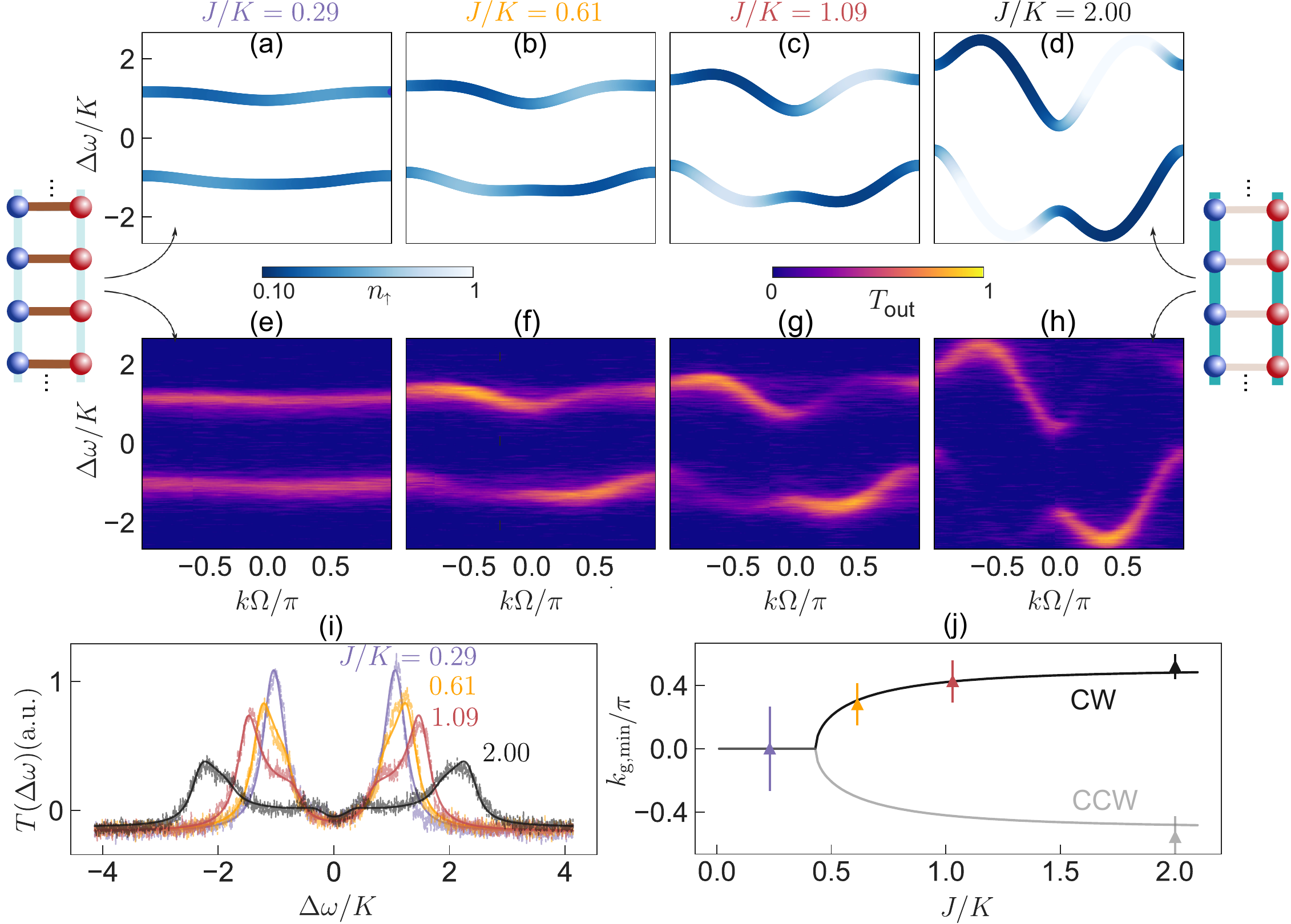}	
\caption{{\bf Observation of phase transition through spin-resolved band structure measurements}. Top row (a)-(d) Theoretical band structure for $\phi_0 = 2.38 \approx 3\pi/4$, for increasing $J/K$. Middle row (e)-(h) Corresponding experimentally measured time-resolved transmission showing good agreement with theory. The ladder insets in the left and right are indicative of the strengths in the pseudospin and frequency axes. $J$ can be continuously tuned by varying the amplitude of the modulation signal. (i) Time-averaged transmission revealing the density of states (DOS). Van Hove singularities due to a diverging DOS are also visible in the transmission, smeared out by the cavity decay rate $\gamma/K = 0.37$. (j) Bifurcation of the energy minimum  in $k$. Data points represent experimentally estimated splittings for band structures shown in the middle row, which agree with the solid lines based on Eq.~\eqref{eq:kgmin}.
}
\label{fig:3_bs_vary}
\end{figure}

The Hall ladder in ultracold atomic systems has been predicted to exhibit a phase transition on increasing $J/K$, from a phase that has a single energy minimum in the ground state (``Meissner" phase) at $k=0$ to a state that has a pair of energy minima at degenerate $k$ points (``vortex" phase)~\cite{hugel_chiral_2014, haug_mesoscopic_2018, atala_observation_2014}. Here we demonstrate a similar transition in the band structure to illustrate the freedom in our system for shaping photonic bands. We adopt the same terminology to facilitate the comparison with existing literature. In our system, $J$ can be easily tuned by changing the modulation voltage while keeping $K$ constant.
For $J/K \ll 1$, the system can be described as a set of decoupled rungs of the ladder. In this regime, the eigenstates are the standing-wave symmetric and antisymmetric supermodes, resulting in flat bands split by $2K$ [Fig. \ref{fig:3_bs_vary}(a), (e) and inset on the left]. Both bands have equal contributions from the CW and CCW legs of the ladder. For $J/K \gg 1$, the two legs of the ladder become decoupled, and we approach the sinusoidal band structure of a 1D-tight binding model with nearest-neighbor coupling~\cite{dutt_experimental_2019-1}. In the intermediate regime, the competition between synthetic SOC and effective magnetic field causes a transition in the band structure from a single minimum at $k=0$ [Fig.~\ref{fig:3_bs_vary}(a), (e)] to two minima [Fig.~\ref{fig:3_bs_vary}(b)-(h)] at~\cite{hugel_chiral_2014}
\be
k_{\rm g, min} = \pm \arcsin \sqrt{\sin^2 \frac{\phi_0}{2} - \frac{K^2 }{J^2 \tan^2(\phi_0/2)}} 
\label{eq:kgmin}
\ee
The experimentally estimated band minima positions agree with the theoretical prediction within measurement uncertainties [Fig.~\ref{fig:3_bs_vary}(j)].

By measuring the time-averaged transmission instead of the time-resolved transmission, we detect the spin-projected density of states (DOS) [Fig.~\ref{fig:3_bs_vary}(i)]. For $J\ll K, \gamma$, two peaks with Lorentzian lineshapes are seen, broadened by the cavity photon decay rate $\gamma$ [Fig.~\ref{fig:3_bs_vary}(i), blue curve]. On increasing $J$, each of these peaks broadens due to the increasing width of the corresponding band structure [orange curve]. Eventually, additional peaks are visible for $J>2\gamma$ [red and black curves], due to van Hove singularities at the edges of both energy bands~\cite{van_hove_occurrence_1953, cortes_photonic_2013, livi_synthetic_2016, kolkowitz_spinorbit-coupled_2017}.

While some aspects such as spin-momentum locking, chiral currents and van Hove singularities have been previously observed in atomic systems ~\cite{mancini_observation_2015, livi_synthetic_2016, kolkowitz_spinorbit-coupled_2017, atala_observation_2014, stuhl_visualizing_2015, cheuk_spin-injection_2012, an_direct_2017, cai_experimental_2019, tai_microscopy_2017}, there are several features that are unique to our photonic implementation. First, we are able to directly measure the dispersion of the \textred{chiral one-way}
modes in synthetic space, thanks to the time-resolved band structure spectroscopy technique, as opposed to mapping of the density-of-states in cold atom experiments~\cite{livi_synthetic_2016, kolkowitz_spinorbit-coupled_2017}. Second, we have access to the entire band structure, including the upper band, which allows us to experimentally observe, for the first time, the chirality switching (Fig.~\ref{fig:4_chiral_current}) when going from the lower to the upper band in a Hall ladder. Finally, our system exhibits frequency conversion, which can have applications in spectral manipulation of light. All of these features are achieved in a simple photonic structure consisting of a single modulated ring, completely based on the synthetic dimension concept.

Future experiments could explore higher-dimensional spin-orbit coupling~\cite{huang_experimental_2016} and topological phases~\cite{lin_three-dimensional_2018} in synthetic space by using multiple coupled rings or additional degrees of freedom such as orbital angular momentum~\cite{luo_quantum_2015, yuan_photonic_2019, schine_synthetic_2016}, temporal multiplexing~\cite{regensburger_paritytime_2012, wimmer_experimental_2017, chalabi_synthetic_2019} and spatial supermodes~\cite{lustig_photonic_2019, maczewsky_extreme_2019}. By incorporating diagonal couplings, additional legs and long-range interactions, phenomena such as chiral Bloch oscillations, bandgap closing and unconventional phases have been predicted recently in Hall ladders~\cite{han_band_2019, zheng_chiral_2017, barbarino_topological_2018, haug_mesoscopic_2018, natu_bosons_2015, ghosh_unconventional_2017, petrescu_precursor_2017, greschner_symmetry-broken_2016, citro_quantum_2018}. The advent of nanophotonic lithium niobate microring modulators with bandwidths exceeding the ring FSR shows promise for realizing synthetic frequency dimensions on chip~\cite{zhang_broadband_2019}. We anticipate that similar synthetic space concepts could be extended to other frequency ranges such as microwaves~\cite{peterson_strong_2019, roushan_chiral_2017, bernier_nonreciprocal_2017}, or to real-space photonic systems where SOC~\cite{vitullo_observation_2017}, chiral quantum emission and spin-momentum locking have been reported~\cite{barik_topological_2018, bliokh_quantum_2015}. Additionally, CW-CCW mode splitting due to point scatterers in microrings have been explored for studying parity-time symmetry~\cite{hodaei_parity-timesymmetric_2014}, non-Hermitian physics~\cite{malzard_bulk_2018-1}, and counterpropagating solitons~\cite{yang_counter-propagating_2017, joshi_counter-rotating_2018}. Simultaneously pumped CW-CCW modes have been also been exploited for Brillouin gyroscopes~\cite{li_microresonator_2017}, light detection and ranging (LIDAR)~\cite{suh_soliton_2018}, spontaneous symmetry breaking~\cite{del_bino_symmetry_2017} and topological insulator lasers~\cite{bandres_topological_2018}.
These ideas can be combined with concepts of gauge potentials, effective magnetic fields and SOC that we have demonstrated here to manipulate and control light in versatile ways.\\

\bibliographystyle{Science}
\bibliography{library_2019_09_01}

\section*{Acknowledgements}
We gratefully acknowledge D.A.B. Miller for initial discussions on the experiment and for providing lab space and equipment.
{\bf Funding:}
This work is supported by a Vannevar Bush Faculty Fellowship (Grant No. N00014-17-1-3030) from the U. S. Department of Defense, and by MURI grants from the U. S. Air Force Office of Scientific Research (Grant No. FA9550-17-1-0002, FA9550-18-1-0379). M.M. acknowledges support from the Swiss National Science Foundation (Grant No. P300P2\_177721).\
{\bf Competing interests:} The authors declare no competing interests. 
{\bf Data availability:} The data related to this study is available in the manuscript and the supplementary materials. Additional data is available from the authors upon reasonable request.

\section*{Supplementary Materials}
Materials and methods\\
Fig. S1\\
References (66, 67)

\section{Materials and Methods}
\subsection{Theory of spin-resolved Floquet band structure from time-resolved transmission}
In this section we prove that the time-resolved transmission through the modulated CW-CCW coupled ring reads out the spin-resolved Floquet band structure of the two-legged Hall ladder.

Following the treatment in Ref.~{\it (8)}, the input-output coupled amplitude equations for modal amplitudes of the system in Fig. 1(b), based on the Hamiltonian in Eq.~(1) are:
\begin{subequations}
	\begin{align}
	\dot a_{m, \cl}(t) &= (im\Omega- \gamma/2) a_{m, \cl} + i\sum_n J_{n-m} (t) a_{n, \cl} + iK e^{im\phi_0} a_{m,\car} + i\sqrt{\gamma_c} \si e^{i\omega t} \\
	\dot a_{m, \car}(t) &= (im\Omega- \gamma/2) a_{m, \car} + i\sum_n J_{n-m} (t) a_{n, \car} + iK e^{-im\phi} a_{m,\cl}\\
	s_{\rm out, \cl}(t) &= i\sqrt{\gamma_c} a_{m, \cl} + \si e^{i\omega t}\\
	s_{\rm out, \car}(t) &= i\sqrt{\gamma_c} a_{m, \car}
	\end{align}
\end{subequations}
where we assume that only the CW mode ($\cl$) is excited from the input waveguide, $\dot a_m \equiv {\rm d}{a_m}/{\rm d}{t}$, and $J_{n-m}(t) = J_{n-m}(t+T)$ is a synthetic dimension coupling introduced by the time-periodic modulation with period $T = 2\pi/\Omega$. $\gamma$ is the total loss rate or the linewidth of each mode, and $\gamma_c$ is the coupling rate of each mode to the bus waveguide. $s_{\rm out, \cl}(t)$ and $s_{\rm out, \car}(t)$ represent the outgoing fields in the forward and backward directions in Fig.~1(a), which couple to the CW and CCW propagation directions within the ring. Defining a gauge transformation
\be 
b'_{m,\cl} = a_{m,\cl} e^{-im(\Omega t+\phi_0/2)-i\omega t}; \ \ \ b'_{m, \car} = a_{m, \car} e^{-im(\Omega t - \phi_0/2)-i\omega t}, \label{eq: bmdefine}
\ee
the coupled amplitude equations in the rotated basis are:
\begin{subequations}
	\begin{align}
	\dot b'_{m, \cl} &= -(i\omega + \gamma/2) b_{m, \cl} + i\sum_p J_{p} (t) b'_{m+p, \cl} e^{ip(\Omega t + \phi_0/2)} + iK  b_{m,\car} + i\sqrt{\gamma_c} \si e^{-im(\Omega t+\phi_0/2)} \\
	\dot b'_{m, \car} &= -(i\omega+\gamma/2) b_{m, \car} +  i\sum_p J_{p} (t) b'_{m+p, \car} e^{ip(\Omega t - \phi_0/2)} + iK b_{m,\cl}
	\end{align}
\end{subequations}
where $p=n-m$. These equations are now in a translationally invariant form along $m$. Hence, we can transform to $k$-space by defining,
\be
\tilde b_{k, \cl}(t) = \sum_m b'_{m, \cl}(t) e^{im\Omega k}; \ \ \ \tilde b_{k, \car}(t) =  \sum_m b_{m, \car}(t) e^{im\Omega k} \label{eq:bk}
\ee
Thus,
\begin{subequations} \label{eq:bkdot}
	\begin{align}
	\dot {\tilde b}_{k, \cl} &= -(i\omega + \gamma/2) \tilde b_{k, \cl} + i\sum_p J_{p}(t) \tilde b_{k, \cl} e^{-ip\Omega k + ip(\Omega t + \phi_0/2)} + iK \tilde b_{k,\car} + i\sqrt{\gamma_c}\, \si\, {T}\, \delta(t+\phi_0/2\Omega-k) \\
	\dot {\tilde b}_{k, \car} &= -(i\omega+\gamma/2) \tilde b_{k, \car} + i\sum_p J_{p} (t) \tilde b_{k, \car} e^{-ip\Omega k + ip(\Omega t - \phi_0/2)} + iK \tilde b_{k,\cl}
	\end{align}
\end{subequations}
Define the column vectors $\ket{{\bf b}_k} = (\tilde b_{k, \cl}, \tilde b_{k, \car})^T$ and $\ket{{\bf s}_{\rm in}} = \si (1, 0)^T$. We can write Eq.~\eqref{eq:bkdot} in a more compact form as,
\be
i \ket{{\bf \dot{b}}_k} = (\omega - i\gamma/2) \ket{{\bf b}_k} +  H_k(t) \ket{{\bf b}_k} -\sqrt{\gamma_c}\, T\,\delta(t + \phi_0/2\Omega-k) \, \ket{{\bf s}_{\rm in}} \label{eq:bkdotvector}
\ee
or,
\be
(\omega + (H_k(t) - i\partial_t) - i\gamma/2) \ket{{\bf b}_k (t)}  = \sqrt{\gamma_c}\, T \, \delta(t+\phi_0/2\Omega-k) \, \ket{{\bf s}_{\rm in}} \label{eq:bkdotfloquet}
\ee
with the $k$-space time-dependent Hamiltonian
\be
H_k(t) = - K \sigma_x - \sum_p J_p(t) e^{-ip\Omega (k-t)} [{\bf 1}_2\cos (p\phi_0/2)  + i\sigma_z \sin(p\phi_0/2)] \label{eq:Hkt}
\ee
As an example, if we choose a single-frequency cosinusoidal modulation, $J_p(t) = J\cos \Omega t$, and apply the rotating-wave approximation (RWA) to keep only the time-independent terms in the Hamiltonian, then the $p=\pm 1$ terms in the summation are the only ones that survive:
\begin{align*}
H_k &= -K\sigma_x - J\sum_p \cos \Omega t\, e^{-ip\Omega (k-t)} \left[ {\bf 1}_2\cos {p\phi_0\over 2}  + i\sigma_z \sin{p\phi_0\over 2} \right] \\
&\overset{\rm RWA}{\approx} - K\sigma_x - \frac{J}{2} e^{-i\Omega k} \left[ {\bf 1}_2\cos {\phi_0\over 2} +  i\sigma_z \sin{\phi_0\over 2} \right] - \frac{J}{2} e^{+i\Omega k} \left[ {\bf 1}_2\cos {-\phi_0\over 2} +  i\sigma_z \sin{-\phi_0\over 2}\right]\\
&= - K\sigma_x - J \left[{\bf 1}_2 \cos k\Omega \cos{\phi_0\over 2} + \sigma_z \sin k\Omega  \sin{\phi_0\over 2}\right]
\end{align*}
which recovers Eq.~(3) in the main text.

Returning to Eq.~\eqref{eq:bkdotvector} for a more general modulation $J_p(t)$ without using the RWA, we note that the Hamiltonian in Eq.~\eqref{eq:Hkt} is time-periodic. Hence, we can define the Floquet eigenstates as,
\be
(- H_k(t) + i\partial_t )\ket{{ \Psi}_{kn} (t)} = \epsilon_{kn} \ket{{\Psi}_{kn}(t)} \label{eq:floquet}
\ee
with $\epsilon_{kn} = \epsilon_k + n\Omega$, and $\ket{{\Psi}_{kn} (t)} = (\psi_{kn,\cl}, \psi_{kn, \car})^T$ being a 2-component column vector representing the CW-pseudospin and CCW-pseudospin components of the Floquet eigenstate. The definition of the Floquet Hamiltonian is of opposite sign from the conventional definition because we use a $e^{+i\omega t}$ frequency convention, which is different from standard quantum mechanics that uses an $e^{-i\epsilon t}$ convention. The inner product in the space of Floquet eigenstates is defined by $\langle \bra{\cdot}\cdot\rangle\rangle_T = (1/T)\int_0^T {\rm d}t \langle \cdot | \cdot \rangle $~\cite{minkov_unidirectional_2018}. In Eq.~\eqref{eq:bkdotfloquet}, we take the inner product with $\bra{{\Psi_{kn}}(t)}$ from the left, to get,
\begin{align}
\left \langle \left \langle {{\Psi}_{kn} (t)} | \left( \omega + (H_k(t) - i\partial_t) + i\gamma/2\right) |{{\bf b}_k (t)}\right \rangle \right \rangle_T &= \left. \sqrt{\gamma_c}\, \langle {\Psi}_{kn} (t)\ket{{\bf s}_{\rm in}} \right|_{t=k-\phi_0/2\Omega} 
\end{align}
Using Eq.~\eqref{eq:floquet} in the above equation, we obtain the expansion coefficients of the intracavity fields $\ket{{\bf b}_k}$ in terms of the Floquet eigenstates $ \ket{ {\Psi}_{kn} (t)}$:
\be
\langle \langle {\Psi}_{kn} (t)\ket{{\bf b}_{k}(t)}\rangle_T = -\left.\frac{\sqrt{\gamma_c} \langle {\Psi}_{kn} (t)\ket{{\bf s}_{\rm in}}} {\omega -\epsilon_{kn} + i\gamma/2} \right|_{t=k-\phi_0/2\Omega} \label{eq:expansion}
\ee
Since $\ket{{\bf b}_{k}(t)}$ is time-periodic, the Floquet eigenstates $ \ket{ {\Psi}_{kn} (t)}$ form a complete basis for expanding them. Thus, we can finally write the output fields as,
\begin{align}
s_{\rm out, \cl}(t) &= i\sqrt{\gamma_c} \sum_m a_{m, \cl} + \si e^{i\omega t}\\
&= i\sqrt{\gamma_c} \sum_m b'_{m, \cl}(t) e^{im(\Omega t+\phi_0/2) + i\omega t} + \si e^{i\omega t}\\
&=  e^{i\omega t}\left [ i\sqrt{\gamma_c}\sum_m b'_{m, \cl}(t) e^{im\Omega (t+\phi_0/2\Omega)} + \si  \right]
\end{align}
From Eq.~\eqref{eq:bk}, we identify the first term in brackets to be $\tilde b_{k, \cl}(t)$ at $k=t+\phi_0/2\Omega$. Using this in combination with Eq.~\eqref{eq:expansion}, we obtain:
\begin{align}
s_{\rm out, \cl}(t) e^{-i\omega t} &= \si + i\sqrt{\gamma_c} \tilde b_{k, \cl}(t) |_{k=t+\phi_0/2\Omega}\\
&= \si + (1, 0) \times \ket{{\bf b}_k(t)} = \si + (1, 0) \times \sum_n \ket{ {\Psi}_{kn} (t)}\,  \langle \langle {\Psi}_{kn} (t)\ket{{\bf b}_{k}(t)}\rangle_T\\
&= \si + i\gamma_c \sum_n \psi_{kn,\cl}(t)\,  \left. \frac{ \langle {\Psi}_{kn} (t)\ket{{\bf s}_{\rm in}}} {\omega -\epsilon_{kn} + i\gamma/2} \right |_{k = t+\phi_0/2\Omega} \label{eq:sout}
\end{align}
As we assumed earlier, only the CW mode is directly excited by the input laser: $\ket{{\bf s}_{\rm in}} = \si (1, 0)^T$. Hence, $\langle {\Psi}_{kn} (t)\ket{{\bf s}_{\rm in}} = \psi_{kn, \cl}^*(t)$. We ignore the first term $\si$ since it is a DC shift in the transmission and we are interested in the time-varying field. Thus, the time-resolved transmission is,
\be
T_{\rm out, \cl}(t)  =\left|\frac{ s_{\rm out, \cl}}{\si} \right|^2 = \gamma_c^2 \left|\sum_n \frac{|\psi_{kn, \cl}(t)|^2}{\omega - \epsilon_k - n\Omega + i\gamma/2}\right|^2_{k = t +\phi_0/2\Omega} \label{eq:tout}
\ee
This shows that the time-resolved transmission through the ring is completely determined by the Floquet quasienergies and eigenstates at a time $t$ shifted by $\phi_0/2\Omega$. This shift originates from the gauge transformation in Eq.~\eqref{eq: bmdefine}.

Apart from a Lorentzian peak at the quasienergies of the system, the transmission is also weighted by the factor $|\psi_{kn, \cl}(t)|^2 = n_{\cl}$, showing the \emph{spin-resolved} nature of the band structure spectroscopy when exciting a single pseudospin (CW, $\cl$) and measuring its transmission. The band structure projected onto the opposite CCW pseudospin ($\car$) can be also measured by setting $\ket{{\bf s}_{\rm in}} = \si (0, 1)^T$ and measuring the CCW transmission (excitation from the right, measurement from the left output in Figs.~1(b) and \ref{fig:S1_setup}).

In the good cavity limit ($\gamma\ll \Omega$), if the synthetic frequency dimension coupling introduced by the modulation is small ($\max|J_p(t)|<\Omega/2$), only the value of $n$ for which the input frequency $\omega$ is closest to $n\Omega$ ($|\omega-\tilde n\Omega| < \Omega/2$) contributes significantly to the summation in Eq.~\eqref{eq:tout}. In this regime, we can further simplify the expression by defining the detuning $\Delta \omega = \omega - \tilde n\Omega$:
\be
T_{\rm out, \cl}(t; \Delta \omega) = \left[ \frac{ \gamma_c^2\, |\psi_{kn, \cl}(t)|^4}{(\Delta\omega -\epsilon_k )^2 + \gamma^2/4}  \right]_{k=t+\phi_0/2}
\ee

The simulated time-resolved transmission in the figures in the main text were calculated from this expression, under the RWA. For calculating the theoretical modal photon numbers $P(m)$ in Fig. 3(f), we take the Fourier transform of Eq.~\eqref{eq:sout}.
%
%

\subsection{Experimental details}
\begin{figure}[!ht]
\centering
\includegraphics[width=\textwidth]{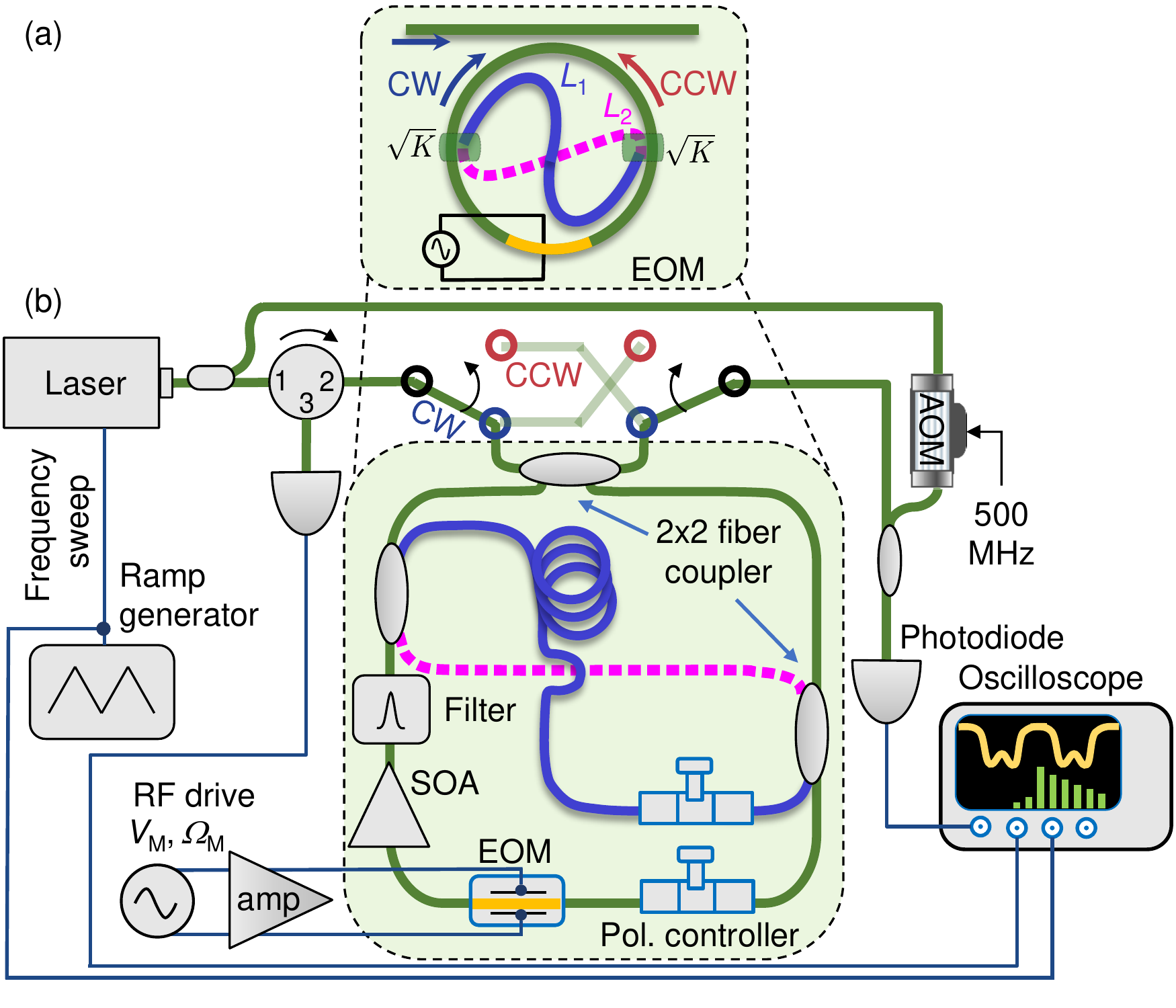}
\caption{{\bf Detailed experimental setup.} The configuration shown here is for exciting and monitoring the CW mode transmission. By switching the fiber connections as indicated by the arrows after the fiber circulator, one can excite the CCW mode (light green fiber paths). Typically we choose $L_1>L_2$. For the band structure measurements (Figs. 2 and 4), we disconnect the AOM path. For the chiral current measurements (Fig. 3), we keep the AOM path connected and use an additional erbium-doped fiber amplifier before the photodiode to boost the signal. The photodiodes have a bandwidth of 5 GHz. SOA: semiconductor optical amplifier. AOM: Acousto-optic modulator, used for frequency shifting the laser by 500 MHz.}
\label{fig:S1_setup}
\end{figure}

The detailed experimental setup is shown in Fig.~\ref{fig:S1_setup}. The photonic cavity consisted of a fiber ring resonator with a length of $L_0 \sim 13.5$ m, corresponding to a free spectral range (FSR) or mode spacing of $\Omega_R/2\pi = 14.8$ MHz.  We used a continuous-wave (cw) narrow linewidth ($<3$ KHz) RIO Orion laser as the input~\cite{numata_performance_2010}. About 90\% of the input laser power was sent through an acousto-optic modulator (AOM) for frequency shifting the laser and used for heterodyne beating with the ring resonator output. The rest 10\% was sent to the fiber ring resonator. The splitting ratios for the various directional couplers were: 99:1 for the ring input-output coupler, 70:30 and 80:20 for the two directional couplers forming the 8-shaped coupler, and 50:50 for combining the AOM output with the cavity output. The EOM was a lithium niobate electro-optic phase modulator with a bandwidth of 5 GHz. To partially compensate for the loss of the EOM ($\sim$3.4 dB) and other component losses, we used a semiconductor optical amplifier (SOA). The amplified spontaneous emission noise introduced by the SOA was filtered by a dense wavelength-division multiplexing filter, which also inhibited spurious lasing in the ring cavity due to higher gain at wavelengths other than the input laser wavelength of 1542.057 nm. Both the ring resonator and the 8-shaped coupler had a polarization controller to align the polarization of the CW and CCW propagating modes to the principle axis of the EOM. We also had a circulator before the ring cavity to monitor the reflected light in the CCW mode upon CW excitation or vice versa. The output was detected with two 5 GHz InGaAs photodiodes. \\

\subsubsection*{Calibration of the frequency axis}
The frequency sweep of the input laser is calibrated using the modulated ring resonator but with no CW-CCW coupling, formed by disconnecting both the connecting waveguides from the directional couplers of the 8-shaped coupler in Fig.~\ref{fig:S1_setup}(b). The modulation frequency is varied till the modulation sidebands overlap with the adjacent resonances, resulting in flattened transmission spectra corresponding to the density of states. More details about the calibration technique can be found in the Supplementary materials of Ref.~{\it (8)}.\\

\subsubsection*{Selection of the 8-shaped coupler's length and modulation frequency}
For realizing a translationally invariant synthetic frequency dimension, we want a fixed coupling between the CW and CCW modes for all mode orders $m$, as evidenced by a constant splitting between the symmetric and antisymmetric supermode frequencies in the unmodulated cavity. Such a constant splitting for all $m$ is not in general achieved for an arbitrary length of the 8-shaped coupler $L_c = L_1 + L_2$,  since the phase introduced by the length of the 8-shaped coupler varies with the mode number $m$ as $\theta_m = \beta(\omega_m) (L_1 + L_2) = 2\pi m L_c/L_0 $. Note that this phase is different from the Peierl's phase $\Delta \phi (\omega_m) = 2 m\phi_0$ associated with the effective magnetic field, which depends on $\Delta L = L_1 - L_2$. However, for a value of $L_c$ that is an integer fraction of the main cavity length $L_0$ ($L_c = L_0/N, \, N \in \mathbb{N}$), the resonances are equally split every $N$ FSRs, as the phase changes by an integer multiple of $2\pi$: $\theta_{N+m} - \theta_m = N\cdot 2\pi L_c/L_0 = 2\pi$. Accordingly, we form the synthetic frequency dimension by coupling modes separated by $N$ FSRs, using a modulation frequency $\Omega = N\Omega_{\rm R}$. In our experiments, we choose $N=2$, corresponding to a modulation frequency $\Omega = 29.6$ MHz.

Hence, the lengths of the two connecting waveguides $L_1$ and $L_2$ are uniquely determined by the   two conditions for (i) matching the 8-shaped coupler length, and (ii) attaining the desired effective magnetic flux per plaquette $\phi_0$:
\be
L_c = L_1 + L_2 = L_0/N
\ee
and 
\be
\phi_0 = \pi \Delta L/L_0 + \pi = \pi + \pi (L_1 - L_2)/L_0
\ee
Thus,
\begin{equation}
L_{1,2} =  \frac{L_0}{2} \left( \frac{1}{N} \pm \left( \frac{\phi_0}{\pi}-1 \right ) \right) \label{eq:L12}
\end{equation}

To accurately satisfy the above equation, we measure the lengths of the fibers forming the connecting waveguides of the 8-shaped coupler using a frequency-domain approach. We first measure the FSR of the ring cavity without the directional couplers of the 8-shaped coupler. By overlapping the modulation frequency of the EOM with the adjacent resonances while sweeping the laser frequency linearly, we obtain an accuracy of 0.2 MHz for the FSR measurement. Next, we add various fiber components such as the directional couplers and polarization controller, one by one, to measure the new FSR, using the same modulation sideband technique. The length of each component is calculated by converting the change in FSR to a change in length using the expression $\delta L = c/n_{\rm g} \times (1/FSR_1 - 1/FSR_2)$, where $FSR_1$ and $FSR_2$ are the FSRs with and without the component(s) in the ring resonator respectively, and $n_{\rm g}\approx 1.46$ is the group index of light in the fiber. Finally, to make up for the remaining lengths of  the $L_1$ and $L_2$ paths to satisfy Eq.~\eqref{eq:L12}, we add extra fiber to each path as required. The frequency-domain approach provides a better measurement accuracy than a direct physical measurement of the fiber component lengths.

\subsubsection*{Chiral current measurements using heterodyne detection}
As outlined in the main text, for measuring the chiral current, heterodyne detection is essential to distinguish the $m>m_{\rm L}$ modes in frequency from the $m<m_{\rm L}$ modes, where $m_{\rm L}$ is the mode closest in frequency to the laser. We use a free-space AOM in a frequency shifting mode by blocking the 0th order diffraction and collecting the 1st order diffraction. Due to the coupling out of the fiber before the AOM and back into fiber after, the frequency conversion efficiency is rather low ($\sim 1\%$), especially due to the high operating frequency of the AOM ($\delta \omega = 500$ MHz). Nevertheless, we offset this low conversion efficiency by using 90\% of the input laser power for the AOM path and by optically amplifying the heterodyne beat signal before photodetection. A large frequency shift $\delta\omega\gg \Omega$ allows us to accommodate $\sim \delta\omega/2\Omega \approx 8$ modes for $m<m_{\rm L}$ before the lower frequency beat notes close to DC start interfering with the measurement. 

For the chiral current measurement, we record a 1-millisecond-long heterodyne interferogram in the time domain at a sampling rate of 8 Gsamples/s while sweeping the input laser frequency. The interferogram is sliced into 4-roundtrip-long time windows with durations of 310 ns. An FFT of each time window yields the mean photon numbers per frequency mode of the excited pseudospin. A 2-point moving average filter was used to reduce noise in the time-domain interferogram.

\end{document}